\documentstyle[aaspp4]{article}

\newcommand{\etal}{{\it et al. \,}}
\newcommand{\kms}{km~s$^{-1}$}

\lefthead{Lee et al.}
\righthead{Dwarf Galaxy UGC 7636}

\begin{document}

\title{Young Star Clusters in the Dwarf Irregular Galaxy, UGC 7636, 
Interacting with the Giant Elliptical Galaxy NGC 4472
}

\author{Myung Gyoon Lee$^1$, Eunhyeuk Kim}
\affil{Department of Astronomy, Seoul National University, Seoul 151-742, 
Korea \\
Electronic mail: mglee@astrog.snu.ac.kr, ekim@astro.snu.ac.kr}

\and

\author{Doug Geisler}
\affil{KPNO/NOAO, Tucson, AZ 85719, USA \\
Electronic mail: dgeisler@noao.edu}

\altaffiltext{1}{Visiting Astronomer, Kitt Peak National Observatory, National
Optical Astronomy Observatories, operated by the Association of 
Universities for Research in Astronomy, Inc., under contract with the
National Science Foundation.
}

\begin{abstract}

We present integrated Washington $CT_1$ photometry of 18 bright blue objects 
discovered in the dwarf galaxy UGC 7636 which is located 
$5'.5$ southeast of the giant elliptical galaxy NGC 4472, the brightest galaxy
in the Virgo cluster. 
Several lines of evidence indicate that UGC 7636 is interacting violently with NGC 4472.
These objects are very blue with colors of $-0.4 < (C-T_1) < 0.6$,  
and their  magnitudes are in the range of $20.6<T_1<22.9$ mag 
which corresponds to
absolute magnitudes of $-10.6< M_{T_1} < -8.3$ mag for a distance modulus
of $(m-M)_0=31.2$.
These objects are  grouped spatially in three regions:
the central region of UGC 7636, the tidal tail region, and
the HI cloud region. No such objects were found in the counter tail region.
It is concluded that these objects are probably young star clusters 
which formed $<10^8$ yr ago
during the interaction between UGC 7636 and NGC 4472.
Surface photometry of UGC 7636 ($r<83''$)
shows that there is a significant excess of blue light along the
tidal tail region compared with other regions. 
The star clusters are bluer than the stellar light in the tidal tail region,
indicating that these clusters might have formed later than most stars in
the tidal tail region which were formed later than most stars in the main body
of the galaxy.

\end{abstract}

\section{INTRODUCTION}

UGC 7636 (VCC 1249) is a dwarf irregular galaxy (Im III-IV: \cite{bin85})
located $5'.5$ southeast 
of the center of the giant elliptical galaxy NGC 4472 (M49), the brightest galaxy
 in the Virgo cluster.
Its proximity to the giant elliptical galaxy (the projected distance between the two galaxies is only 28 kpc) 
and disturbed appearance imply that UGC 7636 might
be experiencing strong interactions with NGC 4472.
Therefore it is an ideal target to study interactions between dwarf
and giant galaxies
 and the evolution of dwarf galaxies in clusters of galaxies.  Basic information
 for UGC 7636 is listed in Table 1.

Surprisingly, there is no HI gas detected in the central region of either UGC 7636 or
NGC 4472. 
Instead, an  $80''\times 40''$ HI cloud of mass $\approx$$7 \times 10^7 M_\odot$
 is found 
between the two galaxies 
(\cite{san87}, \cite{pat92}, \cite{hen93}, \cite{mcn94}).
The structure of the HI cloud is similar, in general, to the optical structure 
of UGC 7636 (\cite{mcn94}).
The radial velocity of the HI cloud is $469\pm3$ \kms (\cite{pat92}, \cite{mcn94}),
which is intermediate between the optical velocities of UGC 7636 
($276\pm78$ \kms: \cite{huc92}) and NGC 4472 ($983\pm10$ \kms: \cite{rc3}).
Recently Irwin \& Sarazin (1996) found in their x-ray study of the NGC 4472 region
a marginally significant X-ray hole in the HI cloud position, suggesting that
the HI cloud may lie at the front side of NGC 4472.

A CO search was carried out for the position of the HI cloud,  with no detection 
(\cite{huc94}, \cite{irw97}).
CO gas of mass $4\times 10^7$ $M_\odot$ has been detected
$2'$ west of the HI cloud ($1'$ south of NGC 4472),
but the measured radial velocity of the CO gas, 883 \kms, indicates that the
CO gas is associated with NGC 4472, rather than with UGC 7636
(\cite{huc88}, \cite{huc94}).

Previous optical imaging studies showed that UGC 7636 has a tidal tail extended
toward NGC 4472 and a counter-tail extended southwest, and that there are 
a few blue knots in the interacting region (\cite{pat92}, \cite{mcn94}).
It has been suggested from these results that the HI cloud may be the gas 
stripped from UGC 7636 via ram pressure and/or tidal interaction 
(\cite{san87}, \cite{pat92}, \cite{mcn94}).  

It is naturally expected that stars and star clusters might have formed 
during the interaction between the two galaxies (\cite{hol96}, \cite{sch96}
 and references therein).
In this paper, we present a study of star clusters in UGC 7636
based on deep CCD images. We have found 18 bright blue star clusters
in UGC 7636, a few of which were previously known (\cite{pat92}, \cite{mcn94}).
The integrated photometry of these bright blue clusters are presented 
as well as the surface photometry of UGC 7636.
The preliminary results presented in a conference proceedings (\cite{kim96})
are superseded by this paper. This paper is organized as follows.
Section 2 explains briefly observations and data reductions, and
Section 3 describes the morphological structure of UGC 7636.
Section 4 presents the photometry of bright blue star clusters in UGC 7636, and
Section 5 shows the surface photometry of UGC 7636.
In Section 6 we discuss the gas stripping and star formation history in UGC 7636. 
Finally  the primary results are summarized in Sec. 7.
  
\section{OBSERVATIONS AND DATA REDUCTION}

Deep Washington system (Canterna 1976)
$CT_1$ images of a $16'.4\times 16'.4$ field centered on NGC 4472 were 
obtained using the $2048\times 2048$ pixel Tektronix T2KB CCD at the prime focus
of the KPNO 4m telescope on the night of 1993 February 26.
The exposure times were $3\times 1000$ s for $T_1$ and $5\times 1000$ s for $C$,
and the seeing was $\approx$$1''.25$.
These images were obtained primarily to study the globular cluster system of NGC 4472,
the results of which are presented in Geisler \etal (1996) and Lee \etal (1997),
 including detailed descriptions of the observations. 

We have chosen a $6'.5 \times 6'.4$ ($811\times 801$ pixels) section 
centered on UGC 7636 in the original field 
to investigate UGC 7636 in detail in this study.
Instrumental magnitudes of the point sources in the images were obtained 
using the digital photometry program DAOPHOT II in 
IRAF\footnote[1]{IRAF is distributed by the National
Optical Astronomy Observatories, which is operated by the Association of 
Universities for Research in Astronomy, Inc., under cooperative agreement
with the National Science Foundation.}.
The resulting instrumental 
magnitudes and colors were calibrated using the standard stars observed
on the same night. The rms scatter of the standard stars was 0.016 and 0.022 for
$T_1$ and $(C-T_1)$, respectively.
The surface photometry of UGC 7636 was  obtained using the ellipse fitting
routine ELLIPSE in STSDAS/IRAF.

We assume UGC 7636 is at the same distance as NGC 4472, since   
the interaction between the two galaxies indicates that they are
close to each other.
We adopt the value of $17.4\pm 1.6$ Mpc for the distance to NGC 4472 which is derived
from the luminosity functions of globular clusters in NGC 4472 (\cite{lee97}). 
At this distance, the projected scale is $1'' = 84.4 $ pc. 
The foreground reddening toward UGC 7636 is known to be negligible, $E(B-V)=0.0$ (\cite{bur82}).

\section{MORPHOLOGICAL STRUCTURE OF UGC 7636}

The details of the structure of UGC 7636 are not easily seen in the original images, 
because it is located in the bright background due to the halo of NGC 4472.
For this reason we have subtracted the contribution due to the unresolved stellar light
of NGC 4472 from the original images using the model galaxy which was created by
modelling the smoothed image of NGC 4472.

Fig. 1 displays a grey scale map of the resulting $C$ CCD image, overlayed by
the intensity contours.
Fig. 1 shows that the inner parts of UGC 7636 are almost circular, but that 
the outer parts are significantly distorted. 
A distorted feature extended toward NGC 4472 in the northwest direction 
represents the tidal tail, and another extension from the center of UGC 7636
 along the southwest direction is a counter tail,  as pointed out
by Patterson \& Thuan (1992) and McNamara \etal (1994).
The detailed structure of the inner region of UGC 7636 is shown in Figs. 2 and 4.
These figures show that there are several bright knots connected by bridges 
in the central region of UGC 7636, and the overall shape of the central region is
like a leading-edge, consistent with the results shown by McNamara \etal (1994).

A contour map of the HI cloud based on the VLA observations by McNamara \etal (1994) is
 also overlayed in Fig. 1.
The peak of the HI cloud ($\alpha_{\rm 1950} = 12^h 27^m 23^s$, 
$\delta_{\rm 1950}=8^\circ 14' 12''$) is $\approx$$2'.2$ (corresponding
to a projected distance of 11 kpc) from the center of UGC 7636, and there is
little HI gas in the main body of UGC 7636.
Comparison of the contour maps of UGC 7636 and the HI cloud shows that the general morphology of the stellar light of UGC 7636 is strikingly similar to
that of the HI cloud, as found by McNamara \etal (1994).

We have created a $(C-T_1)$ color map of UGC 7636 to investigate the spatial
variation of the colors, the contour map of which is overlayed 
on the greyscale map of the $C$ image in Fig. 2.
Fig. 2 shows several interesting features.
First, the colors of UGC 7636 are blue in the central region and 
get redder outward.
Secondly, there is a blue color excess in the tidal tail region,
$\approx$$1'$ (corresponding to a projected distance of 5 kpc) 
northwest of UGC 7636.
Thirdly, there is a group of several blue objects in the HI cloud region,
where these is seen little, if any, stellar light from UGC 7636.
The quantitative measurements of the colors will be given in the following sections.

\section{BRIGHT BLUE STAR CLUSTERS}

We have measured the magnitudes and colors of $\approx$$2,360$ point sources 
in the final images.
Fig. 3 displays the color-magnitude diagrams of the measured objects
in the entire field, the G-region and the F-region.
The G-region corresponds to the main body of UGC 7636, and the F-region
is a control field far from the main body of UGC 7636, as marked in Fig. 1. 
The G-region and F-region are of the same area and are located 
at a similar distance from the center of NGC 4472.

Fig. 3 reveals the presence of several notable features.
The first of these is distinguished by the strong vertical structure with
intermediate colors of $1.0 < (C-T_1) < 2.3$ extending from $T_1\sim 19.5$ mag
to the faintest magnitudes. Most of these are globular clusters in NGC 4472.
These populations are very typical of giant elliptical galaxies 
(\cite{lee93}, \cite{gei96}).
Comparison of the G-region and F-region indicates that there are few, if any,
of these red globular clusters belonging to UGC 7636.

The second dominant population is manifested as the broad horizontal structure
of faint, blue objects with $T_1>23.5$ mag. These are predominantly unresolved
faint background galaxies, as revealed by their blue colors and uniform distribution
over the field (see also \cite{lee93}).

The third population is a small number of very red objects with colors
of $(C-T_1)>2.6$. These objects are distributed uniformly over the field and
are probably  foreground stars in the Galaxy.
Detailed studies of the globular clusters in NGC 4472 and the other two
populations are given in Geisler \etal (1996) and Lee \etal (1997).

The fourth feature is a small number of bright blue objects in the range
$20.6<T_1<22.9$ mag and $-0.4<(C-T_1)<0.6$ 
(represented by the large filled circles in Fig. 3(a)), which are the primary 
subject of this study. 
These objects are marked and identified in the greyscale maps given in Figs. 1 and 4, 
and the photometry of these objects is listed in Table 2. We have used aperture photometry
to measure the magnitudes of some of these objects that were missed in the point-spread
function fitting with DAOPHOT.
We have subtracted the smoothed unresolved stellar light of UGC 7636 from the 
original image (Fig. 1) to show clearly the point sources in Fig. 4.
Figs. 1 and 4 show that these objects are all located in the area associated
with UGC 7636 and are grouped spatially in three regions: 
the central region of UGC 7636 (10 objects), the tidal tail region (3 objects), 
and the HI cloud region (5 objects). 
The positions of these objects correspond to the blue regions
in the color map shown in Fig. 2.
These results indicate that these objects are probably members of UGC 7636.

Some of these blue objects were studied previously.
Patterson \& Thuan (1992) found one blue patch with $(B-I)=-0.1\pm 0.2$ 
at the tip of the tidal tail, which corresponds to C6 in this study.
Later McNamara \etal (1994) presented $UBR$ photometry of several isolated 
blue regions in UGC 7636. The two bluest regions, A and B, in their study correspond to
C2 and C6 in this study, respectively. 

For the purpose of comparison of our colors with previous studies, 
we have estimated the colors of $UBI$  from $(C-T_1)$ colors of these objects
 obtained in this study 
using the transformation relations given by Geisler (1996):
$(U-R)=1.345 (C-T_1)-0.323$, $(B-R)=0.748 (C-T_1)+0.125$, and
$(B-I)=0.989 (C-T_1)+0.191$. 
Table 3 lists the comparison of the three sets
of photometry. The $(B-I)$ color of C6 given by Patterson \& Thuan (1992)
agrees very well with ours, 
but the $(U-R)$ and $(B-R)$ colors given by McNamara \etal (1994) are
significantly redder than ours. 
Considering that our photometry of the globular clusters in NGC 4472
agrees well with that of other globular cluster systems (\cite{gei96}),
the differences between our values and those of McNamara \etal 
are probably due to the uncertainty in the calibration of the photometry 
of McNamara \etal who used the photometry of NGC 4472 (\cite{pel90}) 
for the calibration of their photometry of UGC 7636.

The absolute magnitudes of these objects are derived to be $-10.6<M_{T_1}<-8.3$ mag 
for the adopted distance modulus of $(m-M)_0=31.2$ 
and zero foreground reddening. 
They are as bright as
the bright globular clusters in NGC 4472 seen in Fig.3.
However, they are fainter than the brightest blue star clusters (as bright as
$M_V \sim -15$ mag) found in other interacting
galaxies, the progenitors of which are normal galaxies,
 such as NGC 4038/4039, NGC 3921, NGC 3597, and NGC 6052 (\cite{whi95}, \cite{sch96}, \cite{hol96}).

The color distribution of these objects is weakly bimodal with peaks
at $(C-T_1)=-0.05\pm0.12$ (10 objects, referred to as the BBC group 1 hereafter) 
and $(C-T_1)=0.51\pm0.07$ (8 objects, BBC group 2 hereafter), with a
total mean value of $(C-T_1)=0.19\pm0.30$.
For reference the approximate transformation relation between $(B-V)$ ($(V-I)$) and 
$(C-T_1)$ colors is $(B-V)=0.475(C-T_1)+0.076$ ($(V-I)=0.514(C-T_1)+0.115$)
(\cite{gei96a}). So the corresponding $(B-V)$ colors of the two peaks
are estimated to be $(B-V)\sim 0.0$ and $\sim$0.3 ($(V-I)\sim -0.1$ and $\sim$0.4), respectively.

Considering all the characteristics of these objects found above, 
it is concluded that they are probably star clusters 
associated with UGC 7636.
The blue colors and high luminosities of these clusters suggest that they are 
very young.  
We derive mean ages of $\approx 10^7$ yr 
and $10^8$ yr for the BBC group 1  and the BBC group 2, respectively,
using the $(B-V)$--age relation of a model for the evolution of a single age
population with  solar metallicity given by Bruzual \& Charlot (1993).
Inspection of the structure of these clusters (see Fig. 4) reveals 
that some of them are not as compact as globular clusters, 
but extended or accompanied by faint objects
(which is why we have used aperture photometry to derive their magnitudes). 
This indicates that some of them are loose clusters or OB associations.
We have included  brief notes for the FWHM and morphological features
of these ten clusters in the notes of Table 2. Seven of these ten clusters
belong to the BBC group 1 and the other three belong to the BBC group 2. 

There are several fainter blue objects 
with colors of $(C-T_1)\sim 0$ in the range of $23< T_1 <24$ mag.
Some of these objects may be star clusters of UGC 7636. But it is difficult
to tell whether they are  members of UGC 7636 or background galaxies, because
of the dominant population of background galaxies in the corresponding 
position of the color-magnitude diagram. Indeed, comparison of the F and G region
 CMDs suggests
these fainter blue objects are not associated with UGC 7636.

\section{SURFACE PHOTOMETRY OF UGC 7636}

The surface brightness and color profiles of the entire region centered on UGC 7636
within a radius of $83''$ are displayed in Fig. 5 and are listed in Table 4.

The integrated magnitude and color within $r<83''$ are
$T_1 = 13.80$ mag and $(C-T_1)=1.07$.
Gallagher \& Hunter (1986) presented the integrated colors within the diameter
of $30''$ of UGC 7636, $(B-V)=0.50$ and $(U-B)=0.08$. The $(C-T_1)$ color for
the same region in our photometry is 0.78.  
This value corresponds to $(B-V)=0.45$ and $(U-B)=0.02$,   
only slightly bluer  than the values given by Gallagher \& Hunter (1986).

Fig. 5 shows that the surface brightness profiles of UGC 7636 are well fit
by exponential functions with two components, 
consistent with the results given by Patterson \& Thuan
(1992, 1996) who presented $BI$ surface photometry for the region within
the radius of $80''$ and James (1991) who presented $H$-band data. 

A least-squares fit of the profiles to an exponential law with two components
(for $0''<r<30''$ and $30''<r<80''$)
yields the scale lengths,  
$12''.8$ (= 1.1 kpc) and $28''.1$ (= 2.4 kpc) for the $C$ band and 
$15''.4$ (= 1.3 kpc) and $35''.3$ (= 3.0 kpc) for the $T_1$ band. 
Note that Patterson \& Thuan (1992) obtained the scale lengths 
for the inner region of UGC 7636, $13''.5$ and $18''.3$ for $B$-band and $I$-band,
 respectively, and Patterson \& Thuan (1996)
estimated the scale lengths for the outer region, 
$31''.8$ and $38''.9$ for $B$-band and $I$-band, respectively, which
are consistent with our results: the scale lengths increase with wavelength.
The color of the center of UGC 7636 is as blue as $(C-T_1)=0.7$ 
and the color gets redder 
as the galactocentric radius increases outward, reaching $(C-T_1)=1.4$ at $78\arcsec$,
which is consistent with the result based on $(B-I)$ colors
given by Patterson \& Thuan (1996).

We have divided the entire region centered on UGC 7636 into several sectors 
to investigate the azimuthal variation of the surface brightness and colors.
Fig. 6 displays the surface brightness and color profiles of three sectors
along the tidal tail region (interacting region), the minor axis and
the counter tail region. The position angles of the three azimuthal sectors are
$323^\circ -353^\circ$, $250^\circ - 280^\circ$, and $178^\circ - 208^\circ$,
 respectively.
Fig. 6 shows that there is a significant excess of blue light along the tidal
tail region compared with other regions, becoming the bluest at $r\approx 68''$.
The colors of the counter tail region are much redder than the tidal tail region,
and similar to those of the minor axis region. 

\section{DISCUSSION}

\subsection{Gas Stripping of UGC 7636}

Several lines of evidence suggest strongly that the HI gas has been stripped 
from UGC 7636,
including the location of the HI cloud near UGC 7636, 
the similarity in the structure of UGC 7636 and the HI cloud, 
and the absence of HI gas in UGC 7636.
However, the debate over how the HI gas was removed from UGC 7636 has been controversial.
Two kinds of processes have been proposed to explain the removal of the HI gas
from UGC 7636 (\cite{san87}, \cite{pat92}, \cite{mcn94}).
First, it has been suggested that the gas was stripped out of UGC 7636 by ram pressure 
due to the motion of UGC 7636 through the hot 
dense  halo of X-ray emitting gas (\cite{for85}, \cite{irw96}) in NGC 4472
(see also \cite{gun72}).
Alternatively, tidal interactions between UGC 7636 and NGC 4472 may have removed the gas
as well as some stars from UGC 7636, as typically seen in interacting galaxies.
There are pros and cons for each of these processes in explaining the observational
results of UGC 7636, which are summarized below.

The evidence supporting the ram pressure stripping idea includes:
(a) the location of UGC 7636 in the hot ($T\sim 10^7$ K) and 
dense ($n_0\sim 10^{-3}$ cm$^{-3}$) halo of X-ray emitting gas in NGC 4472,
the brightest galaxy in the Virgo cluster;
(b) the similarity in the structure of UGC 7636 and the HI cloud (\cite{mcn94});
(c) the radial velocity of the HI cloud which is intermediate between those of UGC 7636
and NGC 4472;
(d) the large velocity width (110 \kms) of the spatially extended HI envelope
 (note that the HI line of the cloud consists of both broad and narrow components) 
(\cite{pat92}, \cite{mcn94});
(e) the absence of H$\alpha$ emission in the central region of UGC 7636 
(\cite{gal89});
and
(f) the overall structure of the central region of UGC 7636 which is bow-shaped
(or like a leading-edge) toward the southeast 
along the line connecting UGC 7636 and the HI cloud.

On the other hand, the evidence supporting the tidal interaction scenario includes:
(a) the presence of the faint tidal tail and counter tail of UGC 7636
(\cite{pat92}, \cite{mcn94}, this study);
(b) the blue colors of the tidal tail region and the central region
of UGC 7636 (\cite{pat92}, \cite{mcn94}, this study);
(c) the disturbed morphology of the central region of UGC 7636 
(\cite{mcn94}, this study);
(d) a small velocity width (23 \kms) in the narrow component of the HI line,
which is typical for HI bridges and tails in the interacting galaxies
(\cite{pat92}, \cite{mcn94}); 
and
(e) the estimated value, only $7''$, of the tidal radius of UGC 7636 constrained by NGC 4472, 
indicating the tidal interaction due to NGC 4472 must have reached deep 
into the central region of UGC 7636 (\cite{mcn94}).

The points summarized above suggest 
that both tidal interaction and ram pressure 
have played a role in gas removal and star formation in UGC 7636. 
The similarity in the morphological structure of both
stellar tidal tail and HI cloud implies clearly
that both objects are subject to tidal interactions, 
while the spatial separation of the stellar tidal tail and the HI cloud
indicates that the HI gas was stripped probably by the ram pressure.
McNamara \etal (1994) presented a scenario that
the stars and HI appear to have been tidally distorted in situ, and then the HI
gas was removed from UGC 7636 by the ram pressure, also suggesting that
  only a modest enhancement of star formation appears to have been induced by the interaction.
 They also pointed out that the absence of young stars in the HI cloud
is consistent with the ram pressure stripping. However, this last argument is not 
consistent with the presence of young star clusters in the HI cloud found in our
study. Therefore it is concluded that the UGC 7636 system must have been affected by the tidal interaction before, during and after the gas removal to which ram pressure
might have made a significant contribution. 

\subsection{Star and Cluster Formation History in UGC 7636}

The existence of five young star clusters with ages $<10^8$ yr in the HI cloud
indicates that the HI gas was stripped from UGC 7636 before $\approx$$10^8$ yr ago 
and the star clusters were formed later in the stripped HI gas. 
There is little hint of unresolved stellar light in the HI cloud 
in our deep CCD images. 
The faint unresolved stars, if any, in the HI cloud region must be much fainter 
than those in UGC 7636 and below the detection limit of our images.
Therefore it appears that it is mostly star clusters which were formed recently
and are visible now in the HI cloud. 

The three groups of clusters in the three regions show little difference 
in the mean color (i.e., age) and luminosity, and they all have 
a color distribution which suggest bimodality, as seen in Fig. 3(b). 
This implies that these clusters might have been formed in two different periods 
($\approx$$10^7$ yr and $\approx$$10^8$ yr ago),
and that they were formed in all three regions contemporaneously.

From the presence of bright young star clusters in the three regions and the
similarity of the mean ages of the clusters in each region, we conclude
that there must have been some gas left in UGC 7636 out of which 
star clusters would be formed later, when most of the gas was stripped from UGC 7636.
These bright star clusters were probably formed by tidal interactions, ram pressure,
or thermal pressure due to the hot X-ray gas of NGC 4472.

The mean color of these star clusters is bluer than the color of the stellar 
light of the tidal tail region (Fig. 6), indicating that these clusters were
formed later than most stars in the tidal tail region which were formed later than
most stars in the main body of the galaxy. 

\section{SUMMARY AND CONCLUSIONS}

We have presented integrated Washington $CT_1$ photometry of objects
in UGC 7636 based on deep CCD images.
Surface photometry of UGC 7636 is also presented.
The primary results in this study are summarized  below.

\begin{enumerate}

\item The color-magnitude diagram of the UGC 7636 region shows that there
are 18 bright blue objects which are most likely young star clusters. 
They are grouped spatially in three regions: the central region of UGC 7636,
the tidal tail region and the HI cloud region.
The colors and magnitudes of these clusters
are in the range $-0.4<(C-T_1)<0.6$ and $20.6<T_1<22.9$ mag
($-10.6 <M_{T_1}< -8.3$ mag). 
The blue colors of these clusters indicate they are young, with mean ages
of $10^7$ yr and $10^8$ yr for the two distinct peaks in the color distribution.

\item The surface brightness profiles of UGC 7636 within $r=83''$ 
are well fit by an exponential disk law with two components.
The colors get continuously redder 
outward.
There is a significant excess of blue light in the tidal tail region compared with
other regions.
The colors of the counter tail region are much redder than the tidal tail region,
but similar to those of the minor axis region.

\item Several lines of evidence indicate 
that both tidal interaction and ram pressure have played a role in gas removal
 and star formation in UGC7636.
\end{enumerate}

\acknowledgments

This research is supported in part by the Ministry of Education, Basic
Science Research Institute grant No. BSRI-96-5411 (to MGL).
This research is
supported in part by NASA through grant No. GO-06699.01-95A (to DG) from the 
Space Telescope Science
Institute, which is operated by the Association of Universities for
Research in Astronomy, Inc., under NASA contract NAS5-26555.
Brian McNamara is acknowledged for providing us with the VLA HI map of
UGC 7636.

\begin{deluxetable}{ccc}
\tablecaption{Basic information for UGC 7636.}
\tablenum{1}
\tablehead{
\colhead{Parameter} & \colhead{Information} & \colhead{References} }
\startdata
$\alpha_{1950}$, $\delta_{1950}$ & $12^h 27^m 28^s.2$, +08$^\circ 12' 24''$ & 1 \nl
$l,b$ & 287.1378 deg, 70.1426 deg & 1 \nl
Foreground reddening, $E(B-V)$ & 0 & 2 \nl
Type & Im III-IV & 3 \nl
Optical radial velocity & $276\pm 78$ \kms & 4 \nl
(Optical radial velocity of M49) & $983\pm10$ \kms & 1 \nl
HI radial velocity & $469\pm3$ \kms & 5 \nl
FWHM velocity width of the HI line & 23 and 110 \kms & 6 \nl
$B$-band total magnitudes, $B^T$, $M_B^T$ & 14.94, --16.76 mag &  6 \nl
$I$-band total magnitudes, $I^T$, $M_I^T$ & 12.55, --18.65 mag &  6 \nl
$(B-V)$($D<30''$) & $0.50\pm 0.04$ mag & 7 \nl
$(U-B)$($D<30''$) & $0.08\pm 0.05$ mag & 7 \nl
Size of the HI cloud & $80''\times 40''$ (6.7 kpc $\times$ 3.4 kpc) & 5 \nl
Mass of the HI cloud, M(HI) & $6.9 (\pm 0.4) \times 10^7$ $M_\odot$ & 5 \nl
Number of star clusters & 18 & 8 \nl
Distance modulus, $(m-M)_0$ & $31.2\pm0.2$ mag & 9 \nl
Distance & $17.4 \pm 1.6$ Mpc & 9 \nl
\enddata
\tablerefs{
(1) RC3;
(2) Burstein \& Heiles 1982;
(3) Binggeli \etal 1985, 1993;
(4) Huchra 1992;
(5) McNamara \etal 1994;
(6) Patterson \& Thuan 1992, 1996;
(7) Gallagher \& Hunter 1986;
(8) This study;
(9) Lee \etal 1997.
}
\end{deluxetable}

\begin{deluxetable}{cccccccl}
\tablecaption{Photometry of bright blue star clusters in UGC 7636.}
\tablenum{2}
\tablehead{ \colhead{ID} & \colhead{X[px]} & \colhead{Y[px]} & \colhead{$T_1$}
& \colhead{$\sigma(T_1)$} & \colhead{$(C-T_1)$} & \colhead{$\sigma(C-T_1)$}& \colhead{Remarks}}
\startdata
C1 & 258.52 & 663.34 & 21.04 & 0.04 & {\hspace{-0.2cm}--0.03} & 0.04 & r(aperture) = 8
px \nl
C2 & 267.50 & 606.19 & 21.76 & 0.03 &   0.06 & 0.03 & r(aperture) = 3.5 px \nl
C3 & 285.26 & 602.16 & 22.70 & 0.15 &   0.06 & 0.16 & r(aperture) = 7 px \nl
C4 & 299.05 & 650.47 & 22.14 & 0.02 &   0.58 & 0.03 & PSF mag \nl
C5 & 303.40 & 693.81 & 22.62 & 0.04 & {\hspace{-0.2cm}--0.12} & 0.04 & PSF mag \nl
C6 & 334.29 & 510.38 & 21.89 & 0.07 & {\hspace{-0.2cm}--0.32} & 0.08 & r(aperture) = 8
px \nl
C7 & 335.54 & 471.48 & 22.36 & 0.08 & 0.49 & 0.10 & r(aperture) = 5 px \nl
C8 & 372.26 & 502.82 & 22.27 & 0.14 & {\hspace{-0.2cm}--0.14} & 0.15 & r(aperture) = 7
px \nl
C9 & 393.91 & 346.03 & 22.17 & 0.04 & 0.50 & 0.05 & PSF mag \nl
C10& 399.48 & 385.98 & 20.86 & 0.03 & 0.36 & 0.04 & PSF mag \nl
C11& 408.92 & 354.95 & 21.89 & 0.15  & 0.55 & 0.20 & r(aperture) = 5 px \nl
C12 & 414.48 & 382.30 & 21.19 & 0.09 & {\hspace{-0.2cm}--0.10} & 0.11 & r(aperture) = 4
 px \nl
C13 & 414.58 & 367.83 & 22.63 & 0.05 & 0.01 & 0.06 & PSF mag \nl
C14 & 430.33 & 371.69 & 22.03 & 0.17 &  0.05 & 0.22 &  r(aperture) = 5 px \nl
C15 & 434.16 & 389.86 & 20.59 & 0.05 & 0.53 & 0.06 &  r(aperture) = 5 px \nl
C16 & 434.72 & 403.51 & 22.92 & 0.05 & 0.58 & 0.07 & PSF mag \nl
C17 & 442.68 & 372.07 & 22.93 & 0.08 & 0.46 & 0.09 & PSF mag \nl
C18 & 445.18 & 417.91 & 22.12 & 0.03 & {\hspace{-0.2cm}--0.02} & 0.05 & PSF mag \nl
\enddata
\tablecomments{
(1) C1: FWHM = $2''.0$, elongated structure;
(2) C2: FWHM = $1''.8$, faint companions;
(3) C3: FWHM = $1''.8$, faint background;
(4) C6: FWHM = $1''.5$;
(5) C7: FWHM = $1''.4$;
(6) C8: FWHM = $2''.2$, irregular and elongated structure;
(7) C11: FWHM = $1''.7$;
(8) C12: FWHM = $1''.4$;
(9) C14: FWHM = $2''.2$, extended structure;
(10) C15: FWHM = $2''.4$, extended structure, companions.}
\end{deluxetable}

\begin{deluxetable}{cccccccc}
\tablecaption{Comparison of photometries.}
\tablenum{3}
\tablehead{ \colhead{Object} & \colhead{$(C-T_1)^a$}  &\colhead{$(U-R)^b$}
& \colhead{$(B-R)^b$} & \colhead{$(B-I)^c$} & \colhead{$(U-R)^a_{(C-T_1)}$}
 & \colhead{$(B-R)^a_{(C-T_1)}$}  & \colhead{$(B-I)^a_{(C-T_1)}$} }
\startdata
C2 & 0.06 & 0.25 & 1.12 & \nodata & --0.24 & 0.17 & 0.25 \nl
C6 & --0.32 & 0.10 & 0.82 & --0.1 & --0.75 & --0.11 & --0.13 \nl
\enddata
\tablerefs{$^a$: This study; $^b$: McNamara \etal 1994;
$^c$: Patterson \& Thuan 1992.}
\end{deluxetable}

\begin{deluxetable}{cccccc}
\tablecaption{Surface photometry of UGC 7636.}
\tablenum{4}
\tablehead{ \colhead{$r_{\rm eff}$[arcsec]} & \colhead{$\mu_{T_1}$}
  &\colhead{$\mu(C-T_1)$}
& \colhead{$r_{\rm out}$[arcsec] } & \colhead{$T_1$} & \colhead{$(C-T_1)$}  }
\startdata
  1.36 & $22.29\pm 0.02$ & $0.66\pm 0.03$ &   2.00 & 19.55 & 0.66 \nl
  3.19 & $22.35\pm 0.02$ & $0.70\pm 0.04$ &   4.00 & 18.08 & 0.69 \nl
  5.09 & $22.42\pm 0.04$ & $0.74\pm 0.05$ &   6.00 & 17.25 & 0.72 \nl
  7.03 & $22.53\pm 0.04$ & $0.76\pm 0.06$ &   8.00 & 16.69 & 0.73 \nl
  9.04 & $22.69\pm 0.17$ & $0.78\pm 0.26$ &  10.00 & 16.28 & 0.75 \nl
 10.88 & $22.79\pm 0.19$ & $0.78\pm 0.28$ &  11.72 & 16.00 & 0.75 \nl
 12.63 & $22.90\pm 0.19$ & $0.78\pm 0.30$ &  13.48 & 15.77 & 0.76 \nl
 14.50 & $23.05\pm 0.22$ & $0.85\pm 0.34$ &  15.50 & 15.55 & 0.78 \nl
 16.70 & $23.26\pm 0.13$ & $0.89\pm 0.29$ &  17.82 & 15.35 & 0.80 \nl
 19.19 & $23.46\pm 0.13$ & $0.93\pm 0.15$ &  20.50 & 15.17 & 0.82 \nl
 22.08 & $23.67\pm 0.07$ & $0.97\pm 0.10$ &  23.57 & 15.00 & 0.84 \nl
 25.39 & $23.91\pm 0.24$ & $1.00\pm 0.28$ &  27.11 & 14.84 & 0.86 \nl
 29.22 & $24.16\pm 0.28$ & $1.05\pm 0.31$ &  31.18 & 14.70 & 0.88 \nl
 33.58 & $24.37\pm 0.09$ & $1.06\pm 0.14$ &  35.85 & 14.56 & 0.90 \nl
 38.62 & $24.50\pm 0.19$ & $1.14\pm 0.24$ &  41.23 & 14.42 & 0.93 \nl
 44.37 & $24.68\pm 0.22$ & $1.15\pm 0.28$ &  47.41 & 14.28 & 0.96 \nl
 50.78 & $24.96\pm 0.24$ & $1.09\pm 0.29$ &  54.53 & 14.16 & 0.97 \nl
 58.70 & $25.06\pm 0.20$ & $1.21\pm 0.31$ &  62.71 & 14.02 & 1.00 \nl
 67.53 & $25.28\pm 0.22$ & $1.37\pm 0.35$ &  72.11 & 13.89 & 1.04 \nl
 77.69 & $25.80\pm 0.35$ & $1.41\pm 0.54$ &  82.93 & 13.80 & 1.07 \nl
\enddata

\end{deluxetable}

\newpage

\begin{figure}[1] 
\plotone{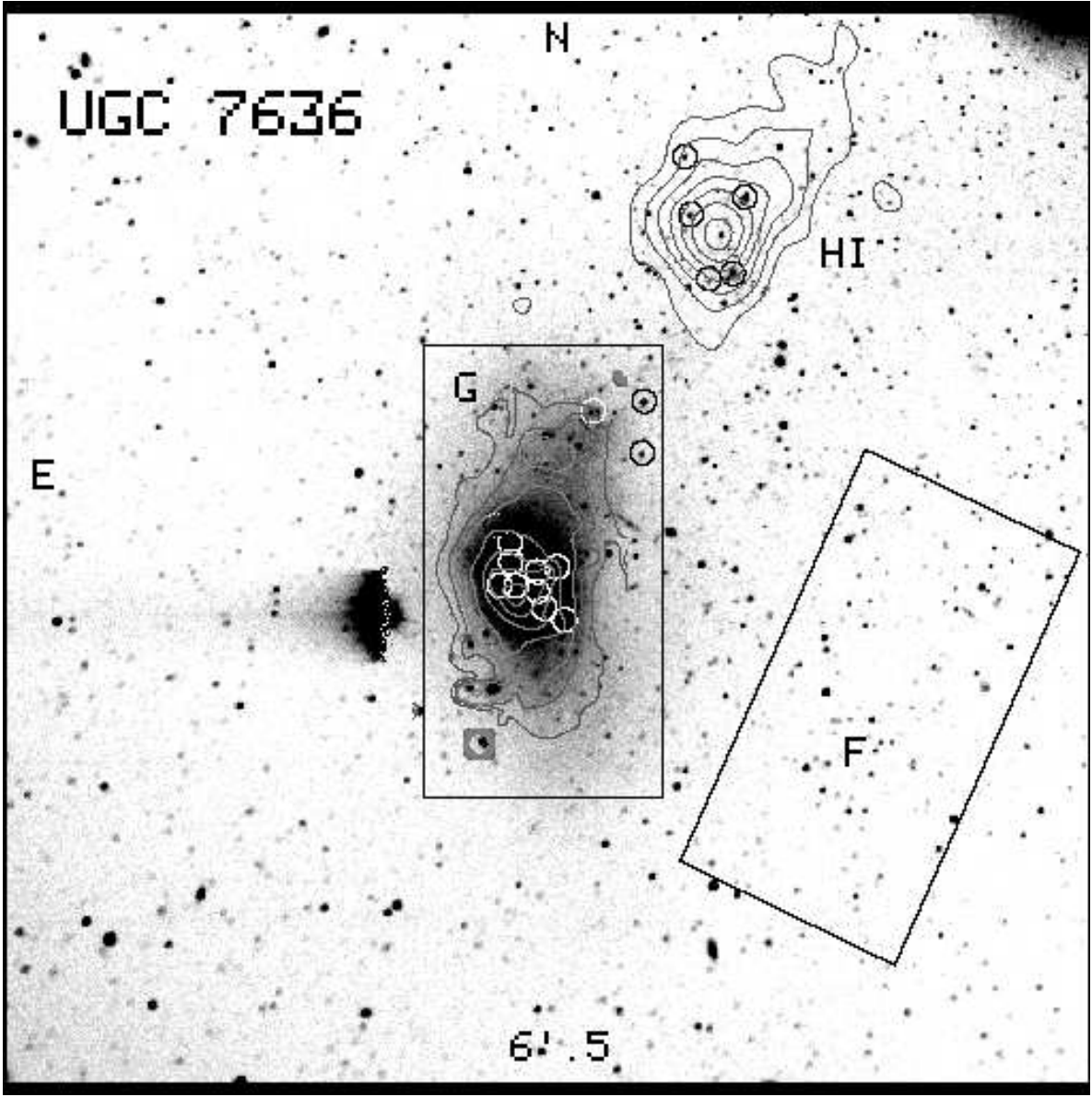}
\figcaption[Lee_figure1.ps]{
A grey scale map of the $C$ CCD image of UGC 7636, overlayed by the intensity
contours.
North is up and east is to the left. The size of the field is $6'.5\times 6'.4$.
NGC 4472 which is off the image is at $5'.5$ northwest of UGC 7636.
The contours labelled as HI represent the VLA HI map given by McNamara \etal (1994).
The dark spot at $\approx$$1'$ east of UGC 7636 represents a saturated
image of a bright foreground star.
The rectangles ($86''\times 163''$) represent the G-region and F-region.
The small circles represent the bright blue star clusters investigated in this study.}
\end{figure}

\begin{figure}[2] 
\plotone{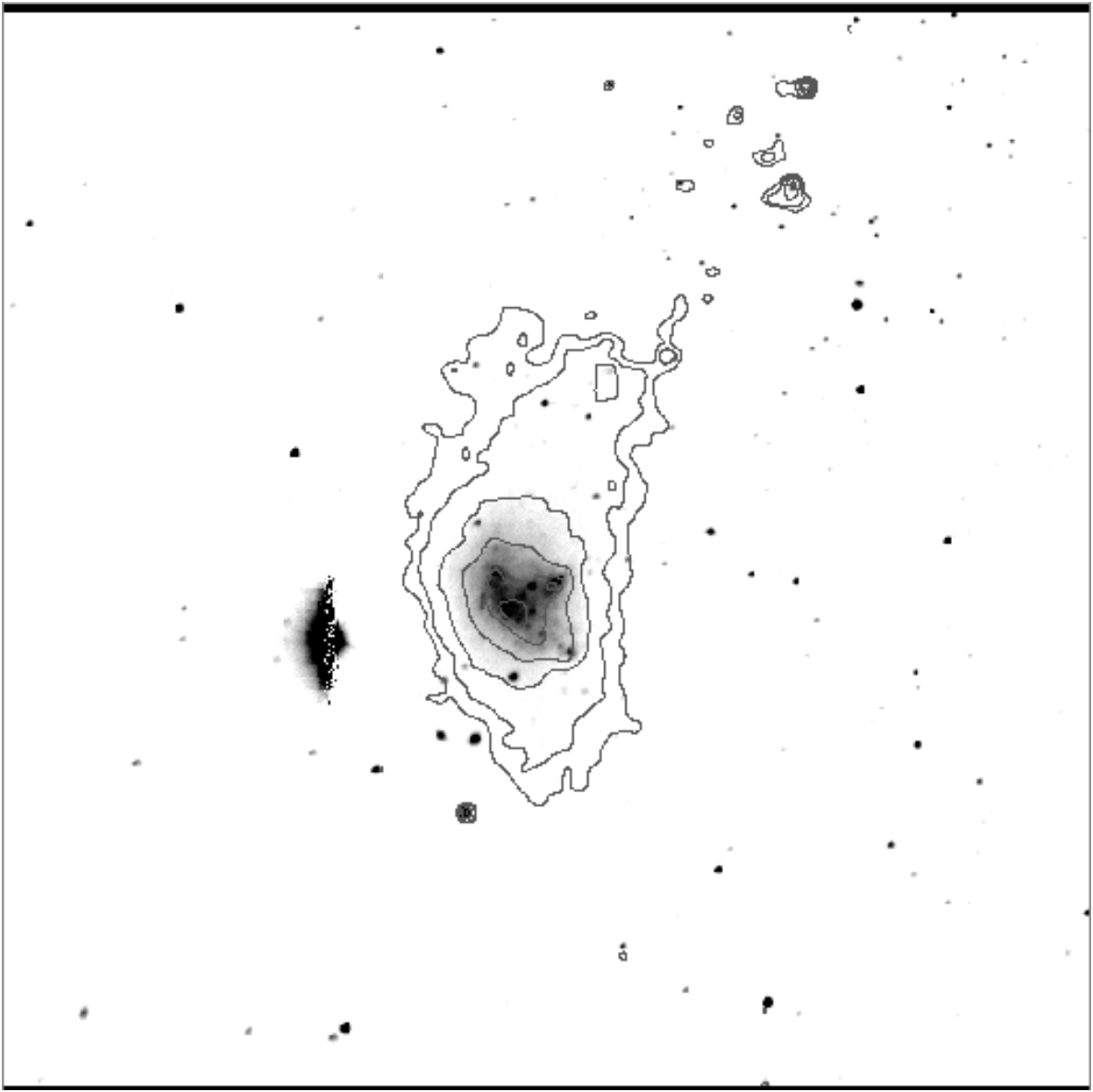}
\figcaption[Lee_figure2.ps]{
A $(C-T_1)$ color contour map of UGC 7636 overlayed on the greyscale map of
the $C$ image.
The size of the field is $4'.9\times 4'.9$.
Contour levels represent approximately the colors of $(C-T_1) =$ 0.71, 0.77, 0.88, 0.99,
1.06, and 1.14,                   
respectively, outward from the center.
}
\end{figure}

\begin{figure}[3] 
\plotone{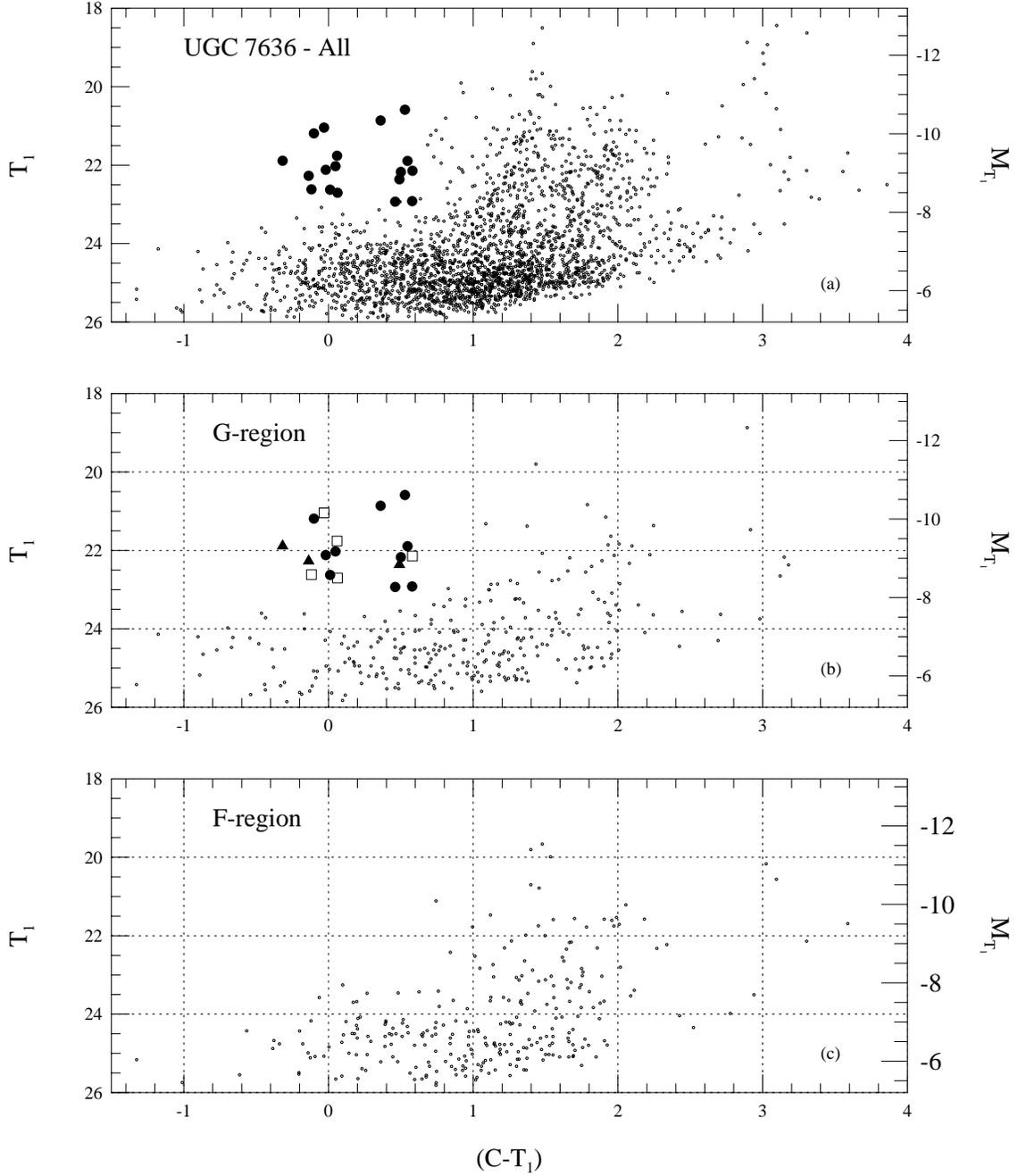}
\figcaption[Lee_figure3.ps]{
(a) $T_1$ vs. $(C-T_1)$ diagram of the measured point sources in the entire field.
Bright blue star clusters are represented by the filled circles and 
the others by the small dots.
(b) $T_1$ vs. $(C-T_1)$ diagram of the measured objects in the G-region.
The filled circles, triangles, and open squares represent the bright blue star 
clusters in the central region of UGC 7636, the tidal tail region 
and the HI cloud region, respectively.
(c) $T_1$ vs. $(C-T_1)$ diagram of the measured objects in the F-region, the control
field.
}
\end{figure}

\begin{figure}[4] 
\plotone{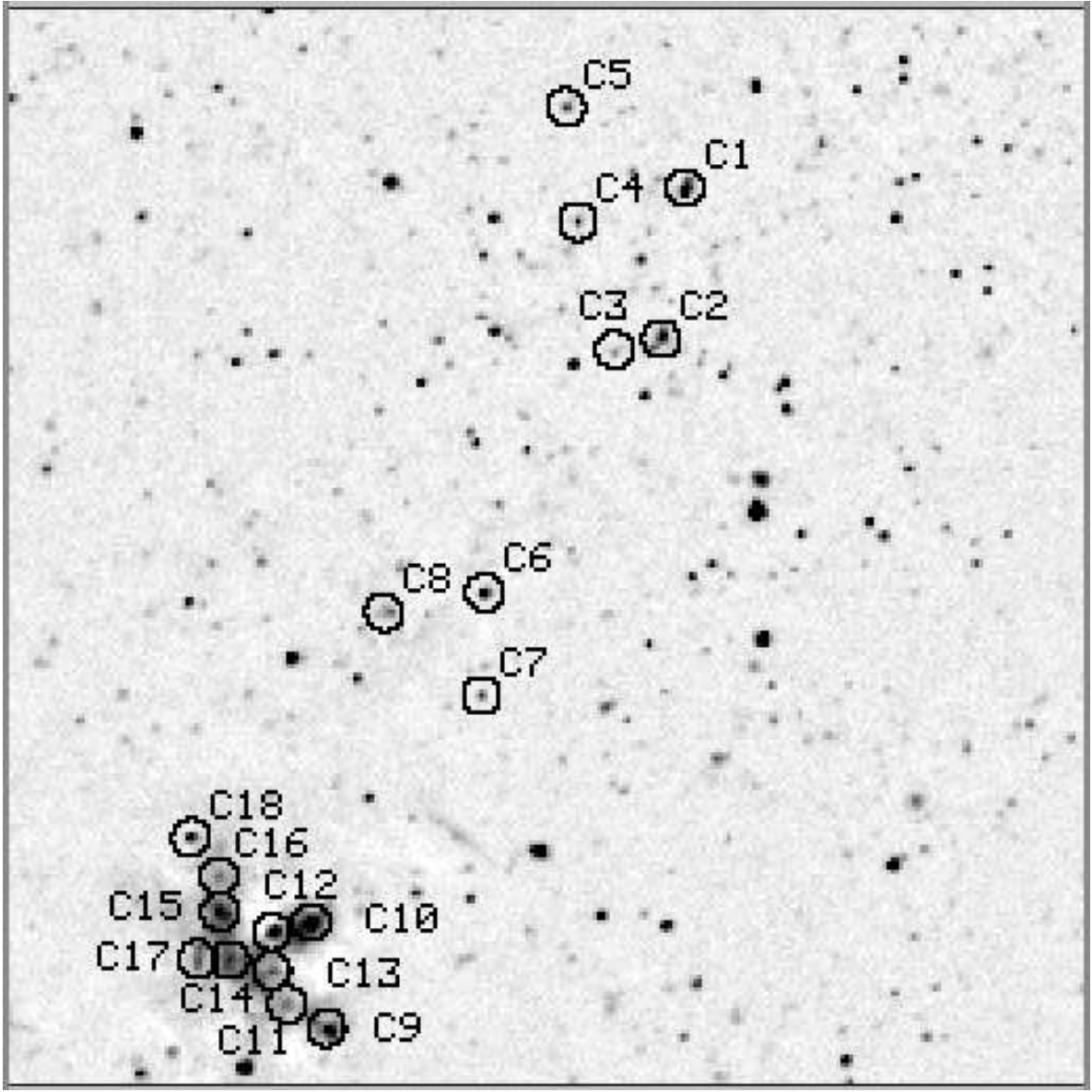}
\figcaption[Lee_figure4.ps]{
Identification of the bright blue star clusters in the $C$ image. 
The size of the field is $3'.25\times 3'.25$.
The smoothed unresolved stellar light of UGC 7636 has been subtracted 
from the original image
to show clearly the resolved point sources.
}
\end{figure}

\begin{figure}[5] 
\plotone{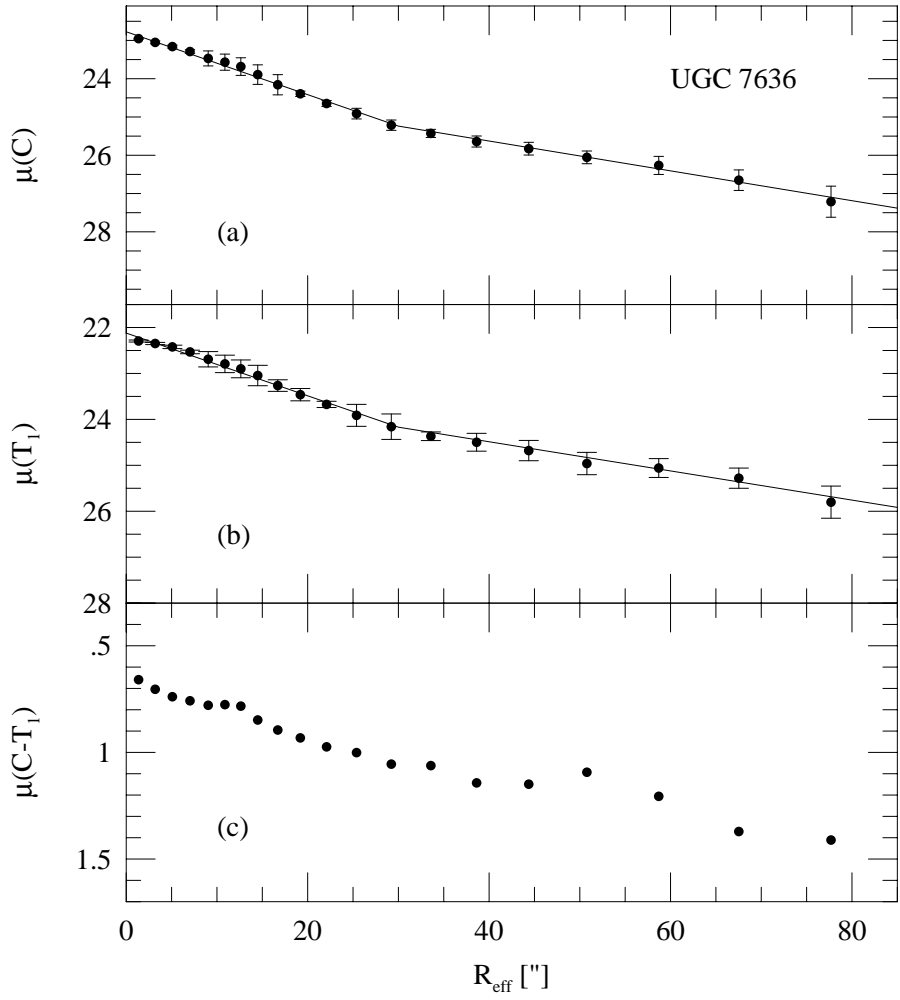}
\figcaption[Lee_figure5.ps]{
Surface brightness and color profiles of UGC 7636. 
The thin solid lines show fits by exponential functions with two components.
}
\end{figure}

\begin{figure}[6] 
\plotone{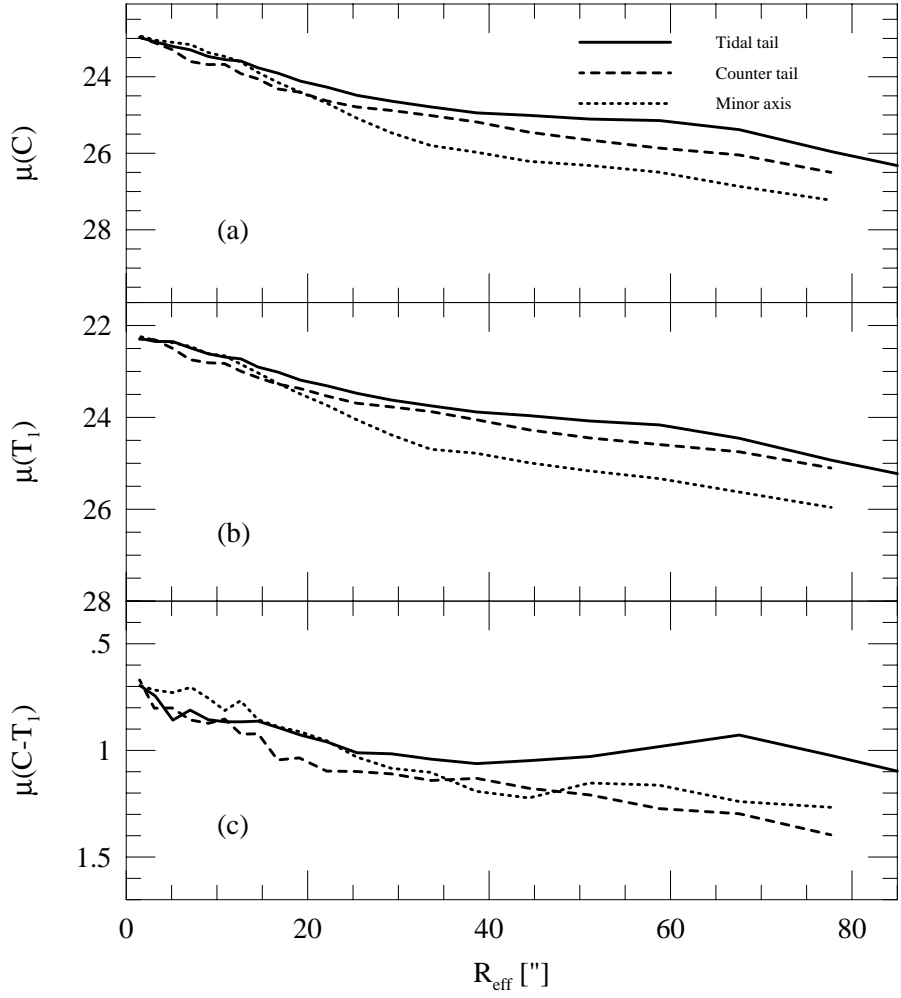}
\figcaption[Lee_figure6.ps]{
Surface brightness and color profiles of three azimuthal sectors of UGC 7636.
The thick solid, dashed, and dotted lines represent, respectively,
the tidal tail, counter tail and minor axis directions.
Note that there is a significant excess of blue light along the direction
of the tidal tail region.
}
\end{figure}

\end{document}